\documentclass[manuscript,nonacm]{acmart}
\usepackage{soul}

\AtBeginDocument{%
  \providecommand\BibTeX{{%
    \normalfont B\kern-0.5em{\scshape i\kern-0.25em b}\kern-0.8em\TeX}}}

\begin{document}

\title[Understanding How Users Collectively Navigate the Complexity of Privacy on Quora]{``Your Privacy is Your Responsibility'': Understanding How Users Collectively Navigate the Complexity of Privacy on Quora}

\author{Varun Shiri}
\email{varun.shiri@polymtl.ca}
\affiliation{%
  \department{Department of Computer and Software Engineering}
  \institution{Polytechnique Montreal}
  \city{Montreal}
  \state{QC}
  \country{Canada}
}
\author{Maggie Xiong}
\email{maggie.xiong@mail.mcgill.ca}
\affiliation{%
  \department{School of Computer Science}
  \institution{McGill University}
  \city{Montreal}
  \state{QC}
  \country{Canada}
}
\author{Jin L.C. Guo}
\email{jin.guo@mcgill.ca}
\affiliation{%
  \department{School of Computer Science}
  \institution{McGill University}
  \city{Montreal}
  \state{QC}
  \country{Canada}
}
\author{Jinghui Cheng}
\email{jinghui.cheng@polymtl.ca}
\affiliation{%
  \department{Department of Computer and Software Engineering}
  \institution{Polytechnique Montreal}
  \city{Montreal}
  \state{QC}
  \country{Canada}
}

\begin{abstract}
\textbf{Abstract:}
In the current technology environment, users are often in a vulnerable position when it comes to protecting their privacy. Previous efforts to promote privacy protection have largely focused on top-down approaches such as regulation and technology design, missing opportunities to understand how to empower users through bottom-up, collective approaches. Our paper addresses this by analyzing what and how privacy-related topics are discussed on Quora. We identified a wide range of interconnected privacy topics brought up by the users, including privacy risks and dangers, protection strategies, organizational practices, and existing laws and regulations. Our results highlight the interplay among the individual, technological, organizational, and societal factors affecting users' privacy attitudes. Moreover, we provide implications for designing community-based tools to better support users' collective efforts in navigating privacy, tools that incorporate users' diverse privacy-related behaviors and preferences, simplify information access and sharing, and connect designers and developers with the user community.
\end{abstract}


\maketitle
\section{Introduction}
The ever-expanding digital footprint of individuals, driven by the pervasive integration of technology into various facets of daily life, raises profound questions about the use and protection of personal information~\cite{zuboff2023age, price2019privacy}. In this technology landscape, an increasing number of users are concerned about their privacy~\cite{gross, Debatin2009, 8835383, XU201142, muipc}. When it comes to privacy protection, however, technology users are in a precarious position: they often have limited options other than those given by the technology developers and service providers~\cite{10.1145/3706598.3713912}, who, in turn, hold conflicting interests between protecting users' privacy and using users' data for improving the product and generating profit~\cite{Mildner2023}.

Previous efforts for promoting user privacy protection were often done from a top-down manner, focusing on approaches such as regulation (e.g., GDPR~\cite{gdpr} and CCPA~\cite{ccpa}), standardization (e.g.,  privacy notices~\cite{Kelley2010} and consent~\cite{Jesus2022}), and technology design (e.g., privacy controls~\cite{Liu2025} and motivational design~\cite{Shiri2024}). While valuable, these efforts tend to have significant shortcomings in practice (e.g., privacy labels not reflecting the actual privacy policy~\cite{ali2024honesty}). They also overlook an important aspect of privacy protection---the bottom-up, collective sense-making and problem-solving process of the technology users when they navigate the complexity of their digital privacy. 

Studies in HCI and CSCW suggested that online social platforms play a significant role in not only connecting users who seek information with those who possess them, but also enabling discursive reflection, resistance, and progression of collective identities, social norms, as well as technological values~\cite{DasS22,DymBFS19,KarizatDEA21}. While existing work has investigated how developers use online forums to ask privacy-related questions and approach privacy issues when developing technology ~\cite{10.1145/3432919, 10.1145/3313831.3376768, 07daf61b7df048c188e185079294a59c, parsons2023understanding}, few studies have explored privacy-related discussions among users, especially how they challenge the social norms around privacy. Those studies either focus on particular aspects of privacy (e.g., permissions for mobile apps~\cite{DBLP:conf/chi/TahaeiAR23}) or users' concerns for certain categories of products (e.g., health tracking apps~\cite{SongMKG24} and smart home devices~\cite{9920029,10179344}). 
Moreover, the predominant narratives in existing work on users' ability and effort to protect their information privacy often portray users as passive and vulnerable~\cite{10.1145/3555154}, with a few exceptions (such as the recent works in the particular context of activism and social movement~\cite{10.1145/3706598.3713870,JIA2022102614}). We argue that the research on information privacy needs more effort on user empowerment rather than protection. This point echos the position by \citet{10596929}, when discussing the research on adolescent online safety.\footnote{\citet{10596929} argues a ``\textit{shift from an authoritarian view of protecting teens to more supportive frameworks that can empower teens to self-regulate and manage online risks meaningfully.}''}

To address this, our research aims to understand how to empower users to collectively navigate the complexity of privacy through effective communication and sense-making. Toward this end, we first have to dissect how the discussion around privacy happens among the users who attend to privacy (e.g., what topics are discussed and how) to understand their attitudes and needs. To achieve this goal, we analyzed privacy-related discussions that happened on a popular Q\&A platform: Quora. Compared to other popular online forums, such as Reddit (e.g., in the study by \citet{10179344}), Quora is a dedicated platform for experience and knowledge sharing and thus provides special advantages for understanding information acquisition and collective sense-making activities. In contrast to developer-centric Q\&A platforms (such as Stack Overflow used by \citet{10.1145/3313831.3376768}), Quora caters to a broader audience, offering valuable insights from a general user's perspective. This platform has an actively growing user base with 400 million monthly unique visitors~\cite{quora_business}. By analyzing a representative sample of 313 Quora questions and their answers, retrieved from a set of privacy-related keywords, we focused on understanding (1) the privacy topics that the users raised questions about and covered in their answers, (2) the type of information sought by the users in privacy-related questions, and (3) the characteristics of the answers to the privacy-related questions.

Our results revealed a wide range of privacy topics discussed by Quora users, touched upon in both questions and their answers. Among the privacy topics, discussions on \textit{privacy protection strategies}, particularly one-stop solutions such as ready-to-use applications or technologies for privacy protection, as well as discussions on \textit{privacy risks and dangers}, particularly societal issues such as surveillance and information leaking, emerged as the most prevalent. We also found that a majority of questions sought \textit{personal opinions} on all aspects of privacy, and many responses were loaded with \textit{emotions} and \textit{personal stories}. Our study highlights the importance of addressing privacy using a more holistic approach that integrates factors across individual, technological, organizational, and societal levels. Our results also provide implications for future community-driven platforms that focus on empowering users in privacy management, considering factors such as diverse user behaviors, different user technical expertise and preferences, ease of access to collective insights, safe discussion spaces, and community engagement with designers and developers. Overall, our work complements previous HCI and CSCW work by providing a bottom-up perspective that informs the design of community-driven technologies that support users to manage their privacy when navigating the increasingly complex technology landscape.

\section{Related Work}
Our work is built on existing literature on how end users perceive privacy, developers' privacy practices and challenges, and studying the online community on Quora. 

\subsection{Investigating End Users' Privacy Attitude}

Different works have investigated end users' opinions, concerns, and behaviors regarding their online information privacy in various contexts. For example, through a survey study, \citet{gross} concluded that users express greater concern towards privacy and security than general computer concerns. Using a similar method, \citet{Elmimouni2024} identified that when compared to privacy researchers and professionals, regular users were aware of a smaller range of technical, conceptual, and practical protection approaches. Through interviews with members of social activist groups, \citet{JIA2022102614} explored the concept of collective privacy and found that the related concerns are constructed and managed through group-specific strategies that emerge from the confluence of internal dynamics, core values, and external relationships.

Various works have also examined users' privacy attitudes and concerns while using different applications and how they perform the delicate risk-benefit analysis when interacting with the app's features. For example, through examining 5,527 security and privacy-relevant reviews from the top 2,583 apps in Google Play, \citet{8835383} identified that users' primary privacy concerns included storage, contact, and location permissions. On the other hand, \citet{Debatin2009} identified that while many Facebook users were aware of and used privacy settings in the application to some extent, they perceived that the benefits of online social networking overpowered the risks. Similarly, \citet{XU201142} found that users value personalization over privacy in the context of Location Aware Marketing, a form of targeted advertising that delivers location-relevant ads.

A few previous studies examined issues raised by users on online forums. For example, the study of \citet{9920029} revealed nine primary areas of privacy concerns from 129,075 privacy-related posts in 66 popular software applications and privacy subreddits, with concerns about ``privacy issues and recommendations'' and ``privacy policies and permissions'' being the most popular. They also observed that privacy narratives (i.e., popular and widespread opinions or recommendations) influence user perceptions about software systems and can be linked to users transitioning from one software to another. Similarly, using the \verb|/r/homeautomation| subreddit as a source, \citet{10179344} observed that users' privacy and security attitudes evolve along the course of their adoption of smart home technologies based on various considerations and contextual factors, such as their technical expertise, preconceptions and assumptions about risks, among others. Focusing on Yahoo Chiebukuro, a Japanese Q\&A site for non-experts, \citet{Hasegawa2022} found that non-expert users primarily sought advice on cyberattacks, authentication, and security software, and had a strong demand for answers related to privacy abuse and account/device management.

While these works investigated end users' privacy concerns expressed in online forums, they primarily focused on categorizing those concerns to inform top-down solutions such as regulations, standards, and technology design. They seldom acknowledge the importance of the collective sense-making and problem-solving activities that those online platforms support. Recent work by ~\citet{10.1145/3706598.3713870} examined users' security and privacy attitudes demonstrated on Twitter, but it is particularly about the period of protests during the ``Justice Pour Nahel'' movement, and concerning the protection of protesters. Our work instead focuses on understanding users' daily online discussion behaviors to inform future community-driven platforms for supporting users' privacy management and protection. Moreover, as previous work suggested, narrative plays a critical part in influencing user's privacy perceptions. While \citet{9920029} studied narratives from a top-down manner (i.e., how events drive privacy discussion as a response), our work uses a bottom-up approach. We delve deep into the discussion thread and reveal how the narratives were formulated by the community (e.g., through the use of personal experience and examples). These unique perspectives and focuses helped us contribute to a deeper understanding of how online communities facilitate collective navigation of the complexity of digital privacy, ultimately informing the design of user-centered community-driven platforms for privacy support.

\subsection{Understanding How Developers Approach Privacy}
Since developers are a major force in shaping how technology is built, how they perceive and practice privacy is of great importance to the overall landscape of online privacy. Online developer forums serve as a community of practice (CoP) for developers to exchange programming-related knowledge and advice with each other. Previous works have explored privacy-related discussions in mainly two online popular developer forums: Stack Overflow~\cite{10.1145/3313831.3376768} and Reddit~\cite{10.1145/3432919}. \citet{10.1145/3313831.3376768} examined 1,733 privacy-related discussions on Stack Overflow, mainly observing that large tech platforms like Apple and Google play a key role in shaping the definition of ``sensitive'' information by laying out requirements for communicating privacy information to users through privacy policies and privacy labels. In a related study, \citet{07daf61b7df048c188e185079294a59c} extracted privacy-related advice present in 119 accepted answers in Stack Overflow to find that many align with Hoepman's \verb|inform|, \verb|hide|, \verb|control|, and \verb|minimize| strategies, focusing on regulatory adherence and information confidentiality~\cite{10.1007/978-3-642-55415-5_38}. \citet{10.1145/3432919} observed that Android developers on the subreddit \verb|/r/androiddev| often show passive attitudes towards privacy, rarely discussing privacy issues concerning personal data, and therefore, advocated for increased transparency and incentives for better privacy practices. Replying on analyzing the distribution of privacy-related keywords, \citet{parsons2023understanding} found that discussions in the \verb|/r/iOSprogramming|, \verb|/r/webdev|, and \verb|/r/androiddev| subreddits mainly focus on app permissions and GDPR compliance, with mostly neutral to positive sentiments exhibited by developers towards privacy issues.

Through directly interviewing developers, many previous work aims to understand developers' privacy practices, in particular their challenges. For example, \citet{hadar2018privacy} interviewed 27 developers who performed software design and identified factors influencing developers' privacy decision-making. They classified those factors into cognitive, organizational, and behavioral, based on Social Cognitive Theory. Their finding indicates that developers experience significant obstacles toward better privacy practice --- on the one hand, they have an incomplete understanding of privacy; on the other hand, they have limited options due to the organizations' privacy policy acting as a powerful force directing developers' privacy practices. Privacy advocates within an organization can help shift the organizational privacy culture. \citet{DBLP:conf/chi/TahaeiFV21} interviewed 12 privacy advocates within the software team and suggested that despite their enthusiasm, they faced many barriers to implementing privacy. Their effort to promote privacy needs to be supported by the management team and a critical mass of other developers. \citet{DBLP:conf/chi/TahaeiAR23}'s work also pointed out developers' challenges when designing the mobile app's permission. Their interview with 19 app developers suggested that while developers made an effort to request permission only when necessary, they faced challenges in understanding the scope of some permissions and were constrained by the third-party libraries' permission requests. Their findings on developers' perspective were compared with end users' perspective, obtained through surveying 309 mobile app users; such a comparison reveals that developers and users are on the same page regarding various aspects related to mobile apps' privacy, such as the importance of privacy and that users should be ultimately responsible for their privacy choices. 

Our study primarily focuses on the end user's perspective on privacy. It complements the previous work on developers' privacy-related concerns by providing insights into how users respond to the existing technology and the privacy practices put forward by the developers.

\subsection{Studies of Online Community on Quora}
Previous works have used Quora to study a variety of topics, such as the linguistic structure of question answerability~\cite{maity_language_2017} and detecting experts through features extracted from online traces~\cite{patil2016detecting}. In the context of social computing, Quora offers a rich space to study online communities' concerns and opinions and how they shape their view on different subjects. For example, \citet{10.1007/978-3-030-34971-4_3} studied the nature and factors that drive anonymous posts on Quora and found that users were more likely to discuss topics such as depression, anxiety, social ties, and personal issues anonymously. \citet{su10051509} used Quora discussions to understand what factors influence public interest in climate change information and how it shapes their opinions, finding that Quora users tend to discuss a divergent range of topics like energy, human and societal issues, and scientific research as opposed to the natural phenomena of climate change. \citet{10.1145/3491102.3517600} used the Bengali  Quora forum to examine how marginalized groups can utilize these platforms for collaborative ``decolonization work,'' to reconstruct their identities and recover from the effects of colonization. 
% \citet{zhao} studies the evolution and interest of health topics relevant to autistic people, observing a broad range of discussed topics relating to diagnosis and treatment of the condition, social challenges, parenting, and education issues. 

Our work uses Quora as a source to understand the online privacy considerations of users through an interchange of knowledge and opinions. Compared with other forums such as Reddit, where the online community is further divided by specific topics, Quora serves as a general question-and-answer platform and can be considered an integrated community and a direct parallel to users as of Stack Overflow to developers. 

\section{Methods}

\subsection{Data Collection}
The data collection and sampling process is carried out using a Python script we developed. We used the \textit{quoras} Python package developed by \citet{10.1145/3406865.3418333} to collect discussions on Quora. Leveraging Selenium \footnote{https://www.selenium.dev/}, an open-source, cross-platform web automation framework, the \textit{quoras} package provides various functions to collect information from Quora, including functions to search Quora posts, topics, or users by keyword. 
Because websites update their code and structure from time to time~\cite{883017}, we reconfigured the package to align with the current Document Object Model (DOM) structure of Quora to accurately retrieve discussions. 

We initially employed a process similar to the previous studies of privacy discussions on Stack Overflow~\cite{10.1145/3313831.3376768, 07daf61b7df048c188e185079294a59c}, searching for the keyword ``\textit{privacy}'' in the question title or tags. However, we observed that using only the keyword \textit{``privacy''} was too broad and returned numerous discussions unrelated to privacy in the context of technologies or software applications. To obtain more relevant questions, we refined our search query to include seven specific terms related to information privacy of technologies or software systems: ``\textit{app privacy}'', ``\textit{data privacy}'', ``\textit{internet privacy}'', ``\textit{mobile privacy}'', ``\textit{online privacy}'', ``\textit{app confidentiality}'', and ``\textit{data confidentiality}''. These terms were created by combining the terms \textit{``privacy''} or \textit{``confidentiality''} with terms related to privacy-sensitive technologies found in our initial exploration. %The term ``confidentiality'' was only paired with the terms ``app'' and ``data'' as the other pairings do not represent commonly used phrases. 
Using these keywords, 1,645 unique questions were retrieved from Quora as of May 2022. To make it feasible for an in-depth qualitative data analysis,  we focused on a representative sample by randomly selecting 313 questions from the retrieved dataset as the target of our analysis; this sample size was calculated with a 95\% confidence level and a 5\% margin of error. Then, full discussions for these questions (i.e., questions and all answers to the questions) were scraped using \textit{quoras}. %From the remaining questions, 65 random discussions were analyzed for an exploratory analysis to formulate the initial codebook.  %It is to be noted that \textit{quoras} function performs the keyword search only within the question.

\subsection{Data Analysis}
The data analysis started with an exploration of how Quora users discussed information privacy issues. For this, the first and the third authors randomly sampled and examined discussions from the remaining dataset, excluding the target sample of 313 discussions mentioned above, to conduct an exploratory qualitative analysis.
Particularly, we followed a bottom-up inductive coding approach, using Atlas.ti\footnote{https://atlasti.com/} and following previous guidelines~\cite{saldana2015coding, denzin2008strategies}. The analysis focused on extracting insights from the following three key elements of the discussions to meet our research objectives of understanding: (1) the prominent privacy topics that were touched upon in the discussions and the questions, (2) the type of information that the questions sought, and (3) the recurring elements in the responses. Starting with 30 discussions, we randomly added new discussions to the analysis until data saturation (i.e., no new topics or themes were found). In the end, 65 discussions were coded during this exploratory phase. During this process, the first and third authors frequently met to discuss the identified themes, resolve discrepancies, and revise and refine the codes. After the initial round of coding, an affinity diagramming~\cite{HOLTZBLATT2017127} activity was conducted to group related topics into higher-level themes. A preliminary codebook was created during this process.  
%The codes were then reviewed with all authors to gather feedback and finalize the initial codebook. 

After the preliminary codebook was created, we conducted a three-round refinement process using the sample of 313 discussions. In each of the first two rounds, the first and second authors independently coded a common set of 30 discussions; %Here, we focused on coding at the subtopic level, rather than the broader code groups identified during affinity diagramming. This facilitated a more granular examination of the themes. Subsequent rounds of coding also employed these more specific codes. 
all four authors then convened to discuss their disagreements and improve the codebook, in hour-long meetings. %Disagreements were defined as instances where one author assigned a code (belonging to a specific topic) to a quotation, while the other author did not. 
Topics with a high frequency of disagreement were paid special attention to in our discussions. %By examining the disagreements and the codebook itself, we refined the codebook for improved clarity.
In the final round of refinement, all four authors independently coded a new sample of 60 discussions (30 discussions per author and each discussion coded by two different authors). Following this round, all authors met again to discuss and resolve any remaining disagreements during two meetings, each lasting 1.5 hours. This iterative process of collaborative coding and resolving disagreements helped pinpoint specific groups in the codebook that caused confusion or required clearer definitions. Based on these discussions, the codebook was finalized by reorganizing and merging some topics, as well as clarifying the scope of specific themes.

Finally, the first two authors applied the final codebook to the entire sample dataset (including reapplying any modified codes on the 90 discussions from the refinement rounds). Among the 313 discussions, 38 were considered irrelevant for one of the following reasons: (1) they addressed privacy or security topics in non-software or digital contexts, such as physical privacy and safety procedures, or (2) the questions did not have any responses for us to analyze. Therefore, the following results were based on the remaining 275 discussions.

\section{Results: Privacy Topics}
\label{sec:ResultsRQ1}
\begin{figure*}
    \centering
    \includegraphics[width=0.8\linewidth]{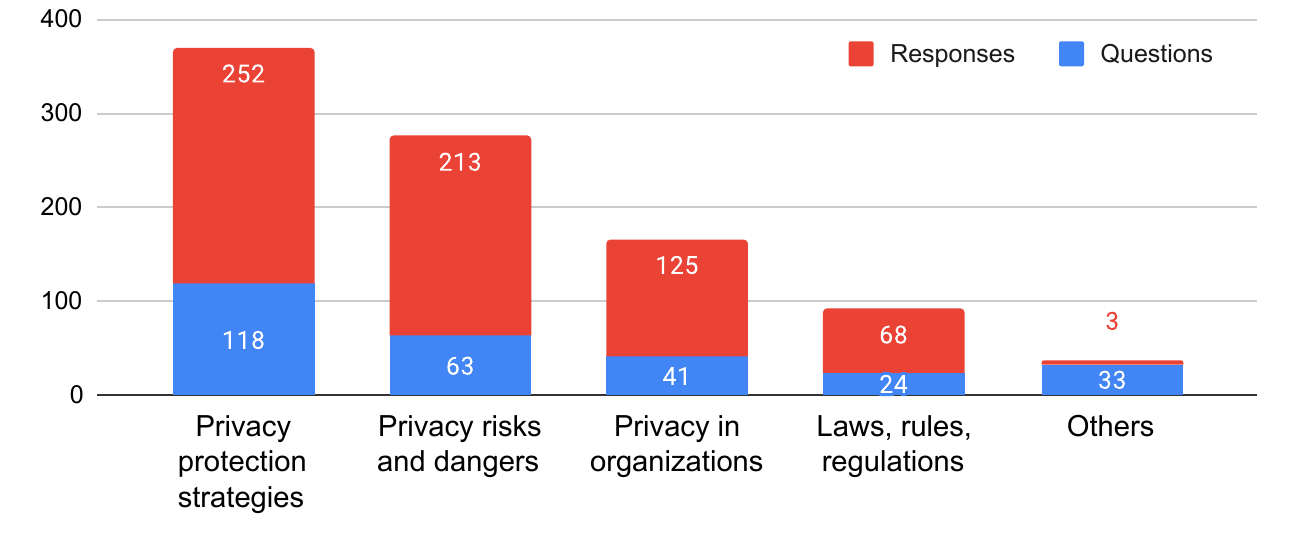}
    \caption{Frequency of privacy topics in questions and responses.}
    \label{fig:risks_summary}
    \Description{}
\end{figure*}

Discussions of privacy on Quora touch on a wide range of topics, mainly concerning (1) privacy protection strategies, (2) privacy risks and dangers, (3) privacy in organizations, and (4) privacy laws, rules, and regulations, either initialized by the questions posted by the Quora users or mentioned in the responses with or without explicitly asking for. Below, we discuss the commonly occurring themes in each category. The numbers in parentheses denote the frequency of the themes, with Q representing the number of their occurrences in questions and R representing the number of occurrences in responses. Each quote can be assigned with more than one theme. Fig. \ref{fig:risks_summary} provides an overview of the distribution of the privacy topics in the questions and responses.

\subsection{Privacy Protection Strategies \texorpdfstring{($Questions=118, Responses=252$)}{Lg}}
\label{subsec:privacy_protecting_strategies}
This topic includes various strategies, such as methods, actions, and plans, discussed by users to protect their privacy. We identified many recurring types of strategies, which are reported below. Because of the complexity of the discussions, each question or response may be coded with more than one type. % along with some other propositions discussed under \textit{Others}.

\textit{\textbf{Applications with Private and Secure Features ($Q=27, R=85$)}}
Many users sought and suggested more privacy-friendly and secure alternatives to applications they used, i.e. apps developed with security best practices and implementing robust privacy and security measures to protect user data and prevent vulnerabilities. Some questions directly ask for such apps, for example, \textit{``What are some best apps or websites to keep my photos with privacy?''}, \textit{``Which are the apps of daily use that actually value user privacy and data?''} and \textit{``What is the best internet browser for protecting privacy on a phone? Why? Do you use it?''}.  In response, users provided recommendations of various apps, for example, when asked about secure messaging apps, a user wrote, \textit{``Telegram is a good choice. It is pretty secure.''}. Secure browsers like Firefox, Brave, and Tor, and privacy-respecting search engines DuckDuckGo were also often suggested, e.g. \textit{``The best browser in terms of privacy is the Mozilla Firefox with DuckDuckGo set as the default search engine.''} Some users also recommended Apple's products and operating systems for enhanced security, as seen in the following response: \textit{``Without a doubt, iOS, iPadOS and macOS are the most private and secure operating systems a common ordinary can use.''}

\textit{\textbf{Applications Built for Privacy and Security ($Q=22, R=60$)}}
Several users also asked for or provided recommendations of applications specifically designed to safeguard individuals' privacy and security while using other apps on their devices. Examples of such questions include, \textit{``What's the best `secure password app' to use to secure internet passwords, credit cards, confidential documents, and inheritance?''} and \textit{``What are the most important internet privacy/security tools for 2019?''} Respondents provided a wide range of recommendations, such as browser plugins and extensions, e.g. \textit{``Where possible, you also want to install a browser plugin called ``Privacy Badger''. It's also a good idea to install something called ``NoScript''. These plugins are easy to get and install in just seconds.''}; privacy protection apps, e.g. \textit{``Leo Privacy Guard helps to protect the private data information of the user like photos, videos personal contacts and many more.''}; and anti-virus tools, e.g. \textit{``Kaspersky Mobile Antivirus: AppLock \& Web Security - Apps on Google Play I just depend on this app's anti-theft feature.''}, among others.

\textit{\textbf{VPN and Firewall ($Q=8, R=71$).}}
Similarly, another common topic found in discussions includes references to VPN and firewall applications to protect an individual's privacy and personal data. Some questions asked for suggestions of good VPNs, e.g. \textit{``What are the best VPN Android apps so far to provide complete privacy?''} Others sought more factual information, posing questions such as: \textit{``How does a VPN work for online privacy?''} and \textit{``Do VPNs guarantee internet privacy?''}. Several responses highlighted the importance and benefits of VPNs and firewalls in protecting users' data: \textit{``There are scenarios where a protected WiFi could also compromise your privacy, make sure your firewalls and other features are updated and use VPN for more security.''} and \textit{``A VPN encrypts internet traffic and routes it through servers located in different locations, effectively hiding your IP address and making it difficult for others to track your online activities.''}. On the other hand, many users also pointed out that using a VPN alone is not sufficient to protect an online user's privacy, as the following statement explains:
\textit{``Earlier only using a VPN or a proxy could easily protect yourself from getting tracked online but nowadays it's just not enough. With a simple proxy or a VPN, it is not possible to prevent browser fingerprinting that can detect even your PC's operating system. The best way to keep yourself untracked and anonymous while browsing online is by using a multi-login anti-detection browser.''}

\textit{\textbf{Anonymity ($Q=2, R=46$).}}
Many respondents recommended providing minimum information to applications and taking measures to stay anonymous online as a means to protect privacy. For example, one user wrote: \textit{``Be cautious when sharing personal information: Don't give out personal information, such as your full name, address, or phone number, to people or websites that you don't trust.''} For certain questions posted by developers,  some users suggested collecting minimal identifiable information, e.g. \textit{``Don't use any identifiable user info. We had a product that didn't use any user info, so we only had their email and proof of payment.''} Two questions asked about solutions that promote anonymity; e.g., \textit{``What are some crazy ideas about what the world needs and should have in terms of online privacy and anonymity to keep companies from spying on us a little too much?''}

\textit{\textbf{Adjusting Settings and Permissions ($Q=2, R=44$).}}
Many responses also refer to changing default settings and selectively granting permissions as a way to safeguard their information privacy. As an example, one user suggested: \textit{``You should ensure that you have enabled any privacy settings to limit what other users can see about you and which information the service can collect...''} While providing advice to protect one's privacy, one user also mentioned that people should \textit{``review every app on your device to make sure that you grant it the minimum permissions it needs to function well, and no more.''} Similarly, two questions inquired about changing an app's settings, e.g., \textit{``Where can I change the privacy settings in the Facebook Messenger app?''}

\textit{\textbf{Encryption ($Q=4, R=39$).}}
Encryption, the process of encoding information or data to make it unreadable, was also a recurring privacy protection strategy mentioned in user questions and responses. Some questions identified were as follows: \textit{``How does the encryption of data impact internet privacy?''}, \textit{``Does the use of encryption ensure online privacy or does it do more harm than good in some cases?''} Several users recommended encryption as a means to secure one's data, e.g., \textit{``Use messaging apps and email providers that offer end-to-end encryption, such as Signal for messaging and ProtonMail for email. These technologies ensure that only the intended recipients can access the content of your messages.''} Encryption was also often mentioned in conjunction with VPNs. For example, one user wrote that a good way to prevent hacker attacks is \textit{``by using a Virtual Private Network (VPN) which will encrypt all network traffic on your device before transmitting them across an open network like the internet. This way everything stays encrypted even after crossing over into an insecure system like the Internet, so no one can get in along the way.''} 

\textit{\textbf{Raising Awareness ($Q=16, R=17$).}}
Various user queries sought clarification on the significance of privacy, asking questions like \textit{``Why is internet privacy so crucial in 2020 and beyond?''}. Others aimed to raise awareness about the significance of privacy, as one user posted, \textit{``How can I convince others that online privacy matters and deserves protection?''} Users also asked for resources to expand their own privacy knowledge, for example, \textit{``What are the best internet privacy blogs?''}. Several responses proposed strategies to heighten privacy awareness, for example, reading the privacy policy, \textit{``Although the privacy policies can be long and complex, they will tell you how the companies maintain accuracy, access, security and control of your personal information and how they use your details. Read carefully before you install them, and close the unnecessary permissions of your apps to avoid that they provide your information to third parties.''} Some users felt that \textit{``it is [the individual's] responsibility to re-assess all your services and devices which you are connected with to see which data you share to find critical privacy lacks.''} \textit{`` Training employees to safely process consumer information''} was another proposed strategy. Another respondent highlighted the importance of teaching the younger population about privacy: \textit{``Chat is the best method to keep youngsters safe online. Continue to teach your youngster about the dangers of the Internet...''}

\textit{\textbf{Caution in Online Behavior ($Q=0, R=29$).}}
Another common recommendation for protecting individuals' online privacy included being vigilant and exercising caution when using the Internet. Users were informed about several key actions to avoid in order to safeguard their privacy. For example, some responses cautioned about clicking unsolicited links: \textit{``Avoid clicking on links from unfamiliar people and hide postings from accounts''} and advised to \textit{``download apps and files from approved sources. Turn off downloads from unknown sources...''} Other respondents suggested regularly clearing data that can be used to track online activity, for example, cookies or cache:  \textit{``Clear your cookies regularly. Most websites may require your cookie to optimize your browsing. However, companies can access your cookies and use your data for targeted ads including location. So remember to clear your cookies regularly especially when you are browsing public websites.''} Similarly, some respondents provided directions to hide personal data, as this comment shows: \textit{``To hide your private data, you won't need any special privacy apps. But you will need selected apps like a file explorer, a file opener etc. Most of the utilities required are already provided in almost all latest file explorers like ES file explorer or any your favourite.''} Another suggestion was to keep social media accounts private and limit the amount of personal information shared, as a user wrote, \textit{``Try not to post personal photos or any pictures of your work, life, families etc. on the Internet''}

\textit{\textbf{Authentication ($Q=1, R=26$).}}
Several participants emphasized the importance of strong authentication techniques for better security. For instance, to manage passwords effectively, respondents recommended using \textit{``a password manager to create and store complex, unique passwords for each online account." } Additionally, users highlighted the benefits of screen lock apps and other secure features like facial recognition or fingerprint ID to safeguard their devices, e.g. \textit{``DroidLock Dynamic Lockscreen... secures the device with a very dynamic, ever changing PIN that virtually no one else except you will ever know} and \textit{``use fingerprinter/Palm Vein Scanner/Face Recognition.''} Users also encouraged the use of additional forms of authentication, like two-factor authentication, because it \textit{``adds an extra layer of security by requiring a second form of verification, such as a unique code sent to your phone and your password.''} One user asked for suggestions for an app to manage authentication: \textit{``What's the best `secure password app' to use to secure internet passwords, credit cards, confidential documents, and inheritance?''}

\textit{\textbf{Software Updates ($Q=0, R=9$).}}
Regularly updating software was one more strategy to maintain privacy, as updates often contain security patches. This is reflected in responses such as  \textit{``Update your OS and apps regularly. Developers add improved security features and remove bugs so keep an eye out for updates.''} and \textit{``Always keep your devices updated! Don't ignore those security fixes and app and operating system updates, they really do matter. Updates are generally free.''}

\textit{\textbf{General Protection Strategies ($Q=28, R=0$).}}
Many users posted general questions related to safeguarding one's privacy. Examples include \textit{``Which layman type practices/ techniques are helping you to secure your Internet Privacy?''},  \textit{``What are the best ways to protect my users' privacy (websites and apps) in 2020?''} and \textit{``What companies or institutions are available to help me with my online privacy management?''} Some users also expressed concerns regarding protecting their privacy against specific apps, including Quora, for example, \textit{``Today, I ordered checks online for a business I run. Now, two hours later, Quora is running an ad in my feed for check printing. How is this done? How can I better protect my online privacy?''} and \textit{``How much data is being captured by search engines like Google, and how should users protect their online privacy?''}

\textit{\textbf{Others ($Q=9, R=8$).}}
In addition to the themes discussed above, a few other infrequent topics emerged from our analysis. For example, some questions were related to protecting the privacy of children: \textit{``What should we do about online privacy protection for children too young to protect themselves?''} Another user inquired about using \textit{``the app privacy report on iOS 15.2''}. One user asked about how individuals in higher positions protect their privacy and security: \textit{``How do CEOs and Ops staff getting IT systems support for their laptops and mobiles and keeping all their data confidential? What IT security tools do they use?''} In the responses, two users appreciated Apple's secure hardware, as one of them wrote, \textit{``T2 Security Chip-enabled Macs become nothing more than a brick until the proper credentials are verified, to unlock it Apple led the industry to require that Safari only accept digital certificates that are not more than 13 months old.''} Some users also suggested more primitive methods to transfer data, \textit{``for example by paper mail, which is more difficult to conduct automatic surveillance on''}, or over the telephone, as one user explained that they \textit{``now do as much financial work and banking as I can over the telephone.''}.

\subsection{Privacy Risks and Dangers \texorpdfstring{($Q=63, R=213$)}{Lg}}
This topic encompasses discussions of potential privacy-related risks or dangers that could result from their privacy attitude or using an application or software. We identified the following prominent types of privacy risks and dangers mentioned by users. Again, one question or response may be coded with multiple types.

\textit{\textbf{Tracking and Surveillance ($Q=13, R=83$).}}
Tracking, collecting, and monitoring individuals' online activities, behaviors, or location data without their explicit consent, often used for targeted advertising or profiling purposes, was one common privacy risk mentioned in responses, as evident in responses such as \textit{``Today we live in a world where everything is monitored online.''} and \textit{``Likewise with virtually all other apps you use, even the seemingly innocent ones like some `flashlight' apps, which are free because they are actually scraping data from your device.''} Many respondents also expressed concerns about the extensive tracking carried out by large companies, for example, \textit{``If you use Google's Chrome browser, a myriad of information is sent back to Google, including a browser fingerprint, your browsing history, detailed geo-location information and details about your device. This information is used to establish your identity on any site that uses Google Analytics (about half the internet, as mentioned above).''} Users also occasionally posted questions about addressing the problem of surveillance, as one user asked, \textit{``What are some crazy ideas about what the world needs and should have in terms of online privacy and anonymity to keep companies from spying on us a little too much?''}

\textit{\textbf{Information Leaking or Hacking ($Q=4, R=75$).}}
Many responses mentioned the dangers of one's personal information being exposed and falling into the wrong hands, where it could be misused by hackers or other malicious parties. For example, when answering a question about why privacy on the web is important, a user wrote: \textit{``Hacking has increased a lot in the past few years. With tech giants knowing everything about their users, it is a high-security concern among online browsers.''} Users also voiced concerns about malware and viruses that could compromise user privacy by stealing personal information or gaining unauthorized access to their devices, as a user expressed: \textit{``Mobile malware has advanced to a new level of sophistication as smart devices continue to gain ground. The number of unique mobile threats grew by 261$\%$ in just two quarters. Increasingly complex malware is taking advantage of a wider range of mobile functionalities to exploit vulnerabilities on the device and in the network.''}. Some questions also sought advice on what to do in the event of an information breach, e.g., \textit{``I found out a confidential data breach in some random company. ... What should have I done?''}

\textit{\textbf{Sale and Misuse of Data ($Q=4, R=57$).}}
On a similar note, responses also mentioned the practice of companies or organizations selling or sharing individuals' personal information without their explicit consent or knowledge, often for marketing or other commercial purposes. For example, one user mentioned: \textit{``...social media is the biggest source of collecting our personal information... They store it on their servers and then sell it to marketers thus, no privacy is provided. I truly think it's a violation of our human rights...''} Another user warned about paying more attention to the terms and conditions, \textit{``You know when you just click accept and quickly move on to the next screen……You just GAVE them the permission to sell your info. They do not HIDE anything, everyone knows Data is more valuable than Oil right now! Please think twice before you click ACCEPT...''} An example of a question on this topic is \textit{``When you use a mobile to open garage doors, does the garage door company monitor your activities and invade your privacy, or at least they can sell your activity data to others if they want?''} Similarly, some responses discussed practices of individuals or entities using an individual's personal data to harass, threaten, or target an individual, e.g. \textit{``You know that I can screengrab that photo in a split second. And then your private photo to a friend is now a digital file that can be shared, copied, turned into memes, and even used in deepfakes to create videos that look like you.''} and \textit{``With issues like social media, big data, cyberbullying and identity theft taking center stage on the news and in lives of people of all ages, internet privacy tools and tactics are in high demand.''} 

\textit{\textbf{Third-Party Data Sharing ($Q=5, R=54$).}}
Many responses also expressed concerns over the common practice of granting third-party entities access to personal data, which may pose privacy risks if the third party does not implement adequate data protection measures. For example, a user wrote \textit{``When some third party has information about you, it grants them power over your action.''} Again, many users underscored that this practice is common in larger companies and organizations, as these comments indicate: \textit{``What's alarming is that social media platforms like Facebook, Instagram and WhatsApp that have access to your personal details can share it with marketers and third party companies.''} and \textit{``If you have an Amazon account, Alexa or Prime, they already got all your data. As to sharing it, if you have started to get tons of targeted ads you data was already shared.''} Two questions asked user's opinions about having their personal data shared, \textit{``Are you concerned about Amazon Sidewalk sharing user information? What implications does this have for online privacy and security?''}

\textit{\textbf{Ignorance About Privacy ($Q=5, R=23$).}}
Meanwhile, some questions highlighted how many users are ignorant about or disinterested in their privacy, making them susceptible to becoming victims of an information privacy breach, such as \textit{``What do you usually tell people who say that online privacy is not an issue?''} Similarly, this topic was discussed in responses to these questions and others, e.g. \textit{``They know that your phone carries with it a LOT of personal information from banking to email to photos, etc. They also know that most people have no idea of how to do information security well or what info security is.''}

\textit{\textbf{Privacy Concerns in Specific Technologies and Apps ($Q=21, R=2$)}}
Several questions inquired about the privacy implications of certain specific technologies, such as artificial intelligence and machine learning, augmented and virtual reality, e.g. \textit{``How is AI and machine learning affecting internet privacy?''} and \textit{``What challenges does AR/VR present in terms of privacy and user experience on mobile platforms?''} Users also expressed privacy concerns for specific products, e.g., \textit{``What are the privacy concerns for TikTok app?''} and \textit{``Should we be worried about internet privacy on Quora?''}

\textit{\textbf{Identification and Deanonymization ($Q=0, R=22$).}}
Another privacy risk mentioned was the possibility of associating anonymous or pseudonymous data with individuals' real identities through different techniques or processes. For example, one user wrote: \textit{``Reddit might not know who you are but people are following you there. And in time, patterns emerge such that people have figured out who is behind those anonymous profiles.''}.

% \textit{\textbf{Identity Theft ($Q=0, R=14$).}}
% Some responses highlighted the risk of unauthorized use of someone's personal information for fraudulent purposes, such as obtaining credit information or making financial transactions in their name. For example: \textit{``When a fraudster obtains access to a variety of scattered, but valuable data that's online, they have the ability to piece them together and develop your identity.''}

\textit{\textbf{Unwarranted Permissions ($Q=0, R=12$).}}
Other responses referred to applications requesting excessive or unnecessary permissions, granting them access to sensitive data or device functionalities beyond what is required for their intended purpose, potentially compromising user privacy. For example, one user gave the following advice to protect one's privacy: \textit{``If you feel the app is asking location unnecessarily, you should deny permission''} 

\textit{\textbf{Lack of Privacy on Internet ($Q=0, R=7$)}}
Some comments also posited that the internet is an open space where privacy is difficult to maintain, as one user wrote \textit{``The sound, reasonable, and pragmatic approach everyone should take to the internet is that you are in public... That's the answer, based on working on it and in it for more than 25 years. The internet is NOT private. Period.''} Users also spoke about the permanency of data posted to the internet, even when deleted, for example, \textit{``You can't delete something from the internet: Whatever you post online, it is there forever. So think carefully before you post anything - images, comments, or even likes.''}

\textit{\textbf{General Privacy Concerns ($Q=9, R=0$)}}
On some occasions, users asked generic and broad questions related to privacy concerns. This includes questions such as \textit{``What is the biggest single threat to internet privacy in the 2020s and why?''} and \textit{``How to fix Internet privacy issues? Why is data privacy an issue?''} Some questions were also related to individuals' privacy concerns in personal matters, for example, \textit{``Why should I be worried about internet privacy if I only have a small amount of money online?''}

\textit{\textbf{Political Risks ($Q=5, R=0$)}}
Some users additionally asked privacy-related questions related to political subjects, such as \textit{``Should internet privacy be an issue for upcoming elections?''} and \textit{``Is sacrificing your online privacy a price worth paying for Brexit?''}

\subsection{Privacy in Organizations \texorpdfstring{($Q=41, R=125$)}{Lg}}
This topic covers discussions concerning how organizations or companies deal with privacy and the privacy aspect of their services. We identified two main themes in this topic.

\textit{\textbf{Company's Privacy Attitudes and Data Handling Practices ($Q=31, R=105$).}}
Many users discussed the attitudes and practices adopted by a company concerning their users' privacy, touching on how companies collect, store, and process user data. These discussions may focus on ways to promote or violate users' privacy. In particular, the data practices of large companies drew a lot of attention, like Google, Facebook, Amazon, and Apple. 
% People asked questions such as \textit{``How does Google sell confidential data and hide from the government? Is it possible?''} and \textit{``Why is Apple obsessed with privacy?''}. 
As an example of negative perception towards company privacy practice, one user wrote: \textit{``Google is big on selling users' information. This means if you are using Google Chrome and you share your location to a website say Facebook, not only are you sharing your location with Facebook but you also are sharing it with Google.''}. On the other hand, some respondents appreciated the stance of certain companies on their users' privacy, for example, \textit{``Apple has identified a market niche they can occupy. They're the ones that care about you and your data. And since Apple doesn't make money via advertising, they can afford to care about your data and your privacy.''}.
    
\textit{\textbf{Privacy Policies and Terms of Service ($Q=10, R=25$).}}
Some discussions touched on users' perceptions or the use of the privacy policy or terms of service of a certain company or application. One subject discussed involved drafting a privacy policy for a business, as well as the content and nature of the policy, reflected in questions such as \textit{``Where do I get privacy policy for my iOS app?''} and \textit{``How can it be explained to users for privacy policy that the app read and retrieve contacts?''} Respondents provided detailed explanations, for example, \textit{``There are many places that you can get a Privacy Policy for your iOS app: lawyers, generators, or drafting it yourself! Hiring a lawyer is the safest and most reliable option, while drafting it yourself leaves you open to the most risk. Using a generator is a `happy medium' that can provide a suitable Privacy Policy for most iOS apps...''} Additionally, users asked questions about the privacy policies of apps and services they used, for example, \textit{``What popular apps or websites have a privacy policy that should concern me and might affect the degree to which I use those sites or apps?''} Similarly, respondents expressed their attitude towards privacy policy as seen in this comment: \textit{``Nobody reads privacy policies before joining a platform like Facebook or Google. They constitute thousands of words of legalese, a construction that literally requires a law degree to parse and begin to understand.''} 

\subsection{Privacy Laws, Rules and Regulations \texorpdfstring{($Q=24, R=68$)}{Lg}} 
This topic includes discussions related to different privacy laws and regulations and their compliance. Two themes were identified.

\textit{\textbf{Requirements Specified by Regulatory Authorities ($Q=21, R=63$).}}
Various discussions referred to privacy-related laws and regulations established by authoritative entities, as well as the compliance of organizations and applications with these laws and regulations. Some users asked about the implications of implementing certain laws, e.g., \textit{``How will the US Congress and Trump's rollback of the FCC's Internet privacy protections affect Internet privacy for users of ISPs in other countries?''} and \textit{``Is internet privacy no more in India after the new law for VPN providers?''} Others raised questions about the differences in laws and regulations across regions, such as, \textit{``Why does the European Union have strict internet privacy laws?''} and \textit{``How does the US's internet privacy compare to that of other countries?''} This aspect was mentioned in responses to other questions as well, for example, when a user asked if complete internet privacy will ever exist, one response highlighted: \textit{``In the EU, they have extensive privacy laws regarding consumer data — they treat it like consumer protection — including the ``Right to be Forgotten,'' where you can request that all the data associated with your account be destroyed.''} Users also discussed the legal requirements for privacy policies: \textit{``The main risks are that privacy laws can be quite specific, and in some places very comprehensive. If you miss something out, your Privacy Policy may not be suitable at all, and may leave you open to fines or regulatory action.''} Often, specific laws or regulations were also mentioned, as seen in comments such as \textit{``The Federal Trade Commission enforces laws such as the Children's Online Privacy Policy Act and the CAN-SPAM act.''} and \textit{``In the EU, the General Data Protection Regulation has created pioneering conditions for individuals to take control over their online data and empower themselves.''}  

\textit{\textbf{Requirements Specified by App Store Platforms ($Q=4, R=7$).}}
Occasionally, users also talked about requirements imposed by App Store platforms like Google and Apple on their developers or users. For instance, one response stated about privacy policies of Android apps: \textit{``It's also important to disclose specific information being collected from your user i.e: login, camera, contacts, etc. Since February 2017 Google enforces a strict privacy policy requirement on apps requesting sensitive permissions and user data.''} Similarly, another answer mentioned that \textit{``Google requires users of Google Analytics to use a privacy policy. When you sign up for Google Analytics, you consent to their terms that state under, you `(...) will have and abide by an appropriate Privacy Policy (...)'''}  Users also clarified doubts they had regarding the requirements, for example, \textit{``Do you have to reveal your use of third party API's, and their own collection of user data, in your app's privacy policy?''} and \textit{``Does the Apple app store require you to have a terms of use and privacy policy for your iOS app?''}

\subsection{Other topics \texorpdfstring{($Q=33, R=3$)}{Lg}}
Some questions and a few responses touched on topics that did not fall into the categories discussed in the previous sections. For example, some questions were of an abstract or philosophical nature. Such questions include: \textit{``When is internet privacy necessary?''}, \textit{``Is internet privacy a human right?''}, \textit{``How to feel about internet privacy?''} and \textit{``What is the maximum level of online privacy right now?''} 

A few users posted questions related to the difference between privacy and security, for example, \textit{``What is the difference between Privacy and Security with reference to a user's data on the Internet?''}, to which one user responded that \textit{``privacy is about controlling who can access an individual's data, while security is about protecting the data from being accessed by unauthorized individuals or entities.''}

Some queries concerned individuals' attitudes towards their privacy, e.g. \textit{``What sensible reason could someone tell me to give up my internet privacy or any of my freedoms?''} and \textit{``Is getting internet privacy worth spending money on? What do you think?''} Similarly, two responses mentioned this subject. One user wrote, \textit{``Media consumers expect streaming services to offer experiences tailored to their watch history and tastes, but they are also more concerned than ever with what data is stored about them and why.''}

There were additional questions that discussed different topics, such as the dark web, e.g., \textit{``What role does the deep web play now and in the future when concerning internet privacy?''}; and political subjects, e.g., \textit{``What is President Trump's position on internet privacy?''}

\subsection{Relationship Between Question Topics and Response Topics}
\begin{figure}[t]
    \centering
    \includegraphics[width=0.8\linewidth]{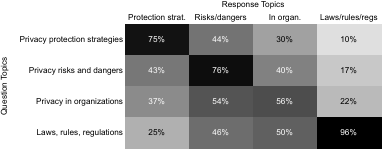}
    \caption{Percentages of questions in each privacy topic answered with responses touching on each topic.}
    \label{fig:relations2}
\end{figure}

To understand how the topics in questions were answered, we analyzed the distribution of privacy topics that appeared in the responses among those raised in the corresponding questions (see Fig.~\ref{fig:relations2}). As expected, privacy topics asked in the questions were most frequently responded to with answers covering the same topics. Interestingly, in questions concerning the topic of privacy in organizations, responses discussed privacy risks and dangers (54\%) almost as frequently as the topic itself (56\%). Privacy laws, rules, and regulations remained the least discussed topic when responding to all question themes except when it was the primary topic asked, in which case, it is almost always mentioned again in the response. Additionally, a noteworthy correlation emerged between privacy risks and dangers and privacy protection strategies; each topic appeared as the second most prevalent topic in responses to questions centered on the other topic.
\section{Results: Types of Questions}
\label{sec:ResultsRQ2}
We identified three categories of information that Quora users sought when posing questions, i.e., opinions, facts, and instructions or guidance. 
%This category discusses the main topics associated with the questions asked in the Quora discussions. The topics may include sub-topics as well depending on the scope of the topic.

\subsection{Opinion \texorpdfstring{($N=135$)}{Lg}}
These questions sought the respondent's opinion or perspective on a privacy-related topic or issue, generally inviting subjective responses. Answering these questions often requires some interpretation and assessment by the respondent based on some evidence. Some questions asked for users' personal or technical opinions on general privacy-related topics, such as \textit{``Do you allow your children to use Facebook's Messenger Kids application given its recent privacy issues?''} and \textit{``Is the best way to protect your privacy on Facebook to not use any applications?''} Other questions were concerned with respondents' interpretation of the intent behind a specific action, policy, or decision of another party or the behavior of individuals in general. Some examples include \textit{``Why will Google like Apple also implement the nutrition label privacy labels on all Android applications distributed on the Play Store?''} and \textit{``With all the talk about online privacy, why do people still choose to put smart devices in their home when companies like Amazon have admitted to listening in to private conversations?''} In addition, some users sought to understand the potential privacy-related consequences or impacts of a specific action, policy, decision, or technology. For example, one user wrote: \textit{``How will stricter internet privacy regulations impact a typical user?''} Another user expressed their concerns about COVID-19 contact tracing apps: \textit{``What are the privacy implications of Google and Apple Covid-19 virus contact tracing apps?''}

\subsection{Factual \texorpdfstring{($N=104$)}{Lg}}
These questions sought factual, objective information about a privacy-related topic or issue. While factual information also requires a certain degree of interpretation, they often have definitive answers, unlike opinion questions, which typically lack a clear right or wrong answer. Some questions in this category asked for technical details about privacy-related problems, such as \textit{``Which web application attack is more likely to extract privacy data elements out of a database?''} and \textit{``Can another application or browser use our privacy and data on a large scale with the help of Chrome?''} Some other questions sought information about privacy laws, regulations or organizational rules, such as \textit{``Does the Apple app store require you to have a terms of use and privacy policy for your iOS app?''} and \textit{``How long before “the right to be forgotten” (internet privacy) becomes law in the United States?''} Moreover, users also posed questions to gain factual information about certain applications or technologies, such as \textit{``What are some of the privacy features on the Signal app?''} and \textit{``Is the cloud a safe environment to store confidential data?''}

\subsection{Instruction or Guidance \texorpdfstring{($N=66$)}{Lg}}
These questions specifically looked for information or guidance on performing a particular action. The instruction provided can be subjective or objective. The responses to these questions generally provide a basis for action for the questioner. For example, \textit{``How do I maintain privacy of user data in SAAS application?''} and \textit{``Where can I get some templates/drafts of terms of services and privacy policies for my web application?''} Some users also asked for suggestions of tools that can help users protect their privacy, such as \textit{``What are the applications one should install before going online to ensure privacy and protection?''} and \textit{``What is the best internet browser for protecting privacy on a phone? Why? Do you use it?''}

\subsection{Relationship Between Question Types and Question Topics}
\begin{figure}[t]
    \centering
    \includegraphics[width=0.6\linewidth]{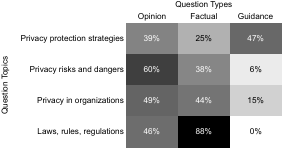}
    \caption{Percentages of questions in each privacy topic that were asked with each question type.}
    \label{fig:relations1}
\end{figure}
We analyzed the proportion of the three question types for each privacy topic covered in the questions to understand the aspects that the users were interested in for each topic (see Fig. \ref{fig:relations1}). The percentages were calculated as a ratio of the total number of times a privacy topic appeared in the responses for a corresponding privacy topic in the question against the total number of occurrences of that privacy topic in the questions. For questions about privacy risks and dangers and privacy within organizations, users most frequently searched for others' opinions (60\% and 49\% respectively). Meanwhile, suggestions or guidance were least sought for all topics except questions on privacy protection strategies (47\%). Further, most of the questions about privacy laws and regulations were factual questions (88\%), with about half asking for opinions (46\%) and no inquiries requesting guidance on legal matters.

\section{Results: Characteristics of Responses}
\label{sec:ResultsRQ3}
To understand how the community reacts to the privacy questions, we characterize the responses based on their style, nature or format. We observe that the responses are rich in strong sentiment, concrete and illustrative examples, external resources, and personal stories.

\subsection{Emotional Arousal \texorpdfstring{($N=147$)}{Lg}}
Users sometimes expressed strong emotions in their responses that hinted at their perception and attitude toward the subject being discussed. Some users conveyed \textbf{contempt} towards certain matters or to the questioner, as seen in this statement: \textit{``Hello naive stranger! Companies do not SELL data to the government in China. Companies provide data access to law enforcement when requested. There is no money involved.''} \textbf{Sarcasm} was a tactic sometimes employed when responding to questions that the respondent viewed as seemingly trivial; e.g., \textit{``I have a very simple test for how important online privacy should be to you: Please post your full name, address, phone number, social security number, bank account numbers, and credit cards. Post the names, ages, and schools attended by your children (this is of course after posting your bank account information). Please put up an active webcam in your bathroom pointing at your toilet. Once you have done all of the above, let's continue this discussion about online privacy.''} 

Occasionally, users also indicated a sense of \textbf{helplessness} towards their privacy in responses such as this one: \textit{``But reflect, what are they going to do with this information? You really cannot do much about that either. So why worry about this, all you will do is put more worry in your head?''} 
Similarly, some users added depth to their answers by including \textbf{thought-provoking questions} that encourage readers to reflect on the topic.; e.g. \textit{``When someone hacks and steals data from any platform, be it FB, google, yahoo, a bitcoin exchange, who can say confidently that their information has not been harvested, and by extension, that they have been compromised? In this, the twenty first century, can we do without all the online services? ...Can we accept that if and when we use these services, we tend to loose our identity a little at a time? I think, a resounding yes. Is it possible that every netizen, (internet user) has somewhere, somehow been compromised, and we are totally unaware of it? I think, yes. Solutions, any one?''} In addition, some users expressed \textbf{radically different opinions } compared to other responses to the same or similar question. To illustrate, when a user asked about the most secure operating software for phones, one of the responses was, \textit{``In my opinion, online privacy regulation should be mandatory and enforced by draconian punishments. I should be the final arbiter of where my data goes, who can see it, what it can be used for.''} 

\subsection{Example \texorpdfstring{($N=136$)}{Lg}}
Providing examples or illustrations to support a point or idea was a common tactic used in responses. For instance, one user provided the following account while explaining the need to care for one's privacy: \textit{``Earlier this year, a young woman purchased a few nondescript items such as cotton balls, unscented lotion and some vitamins. Based on what the company already knew about her, they were able to correctly predict that she was pregnant, and began targeting her for baby items by sending her coupons in the mail. The issue? She was a teenage girl, and these coupons alerted her father that she was indeed pregnant.''} Another user included a diagram to explain the concept of VPN: \textit{``Here's a quick drawing to make this more clear, I hope. The traffic between your device and the bank is represented as the pink arrows. It will be protected by the VPN, but only between you and your VPN provider. From your VPN provider to your bank it's relying on the TLS encryption that's inherent to HTTPS.''} Many users also employed illustrative comparisons to clarify their point. For example, one user wrote about the risk of having one's online privacy compromised: \textit{``Anytime you go online you take that risk. Just as you do everytime you get in a car or buy a lottery ticket. The difference though is i still have a choice to get into a car or buy a lottery ticket but my choice to use the internet is being is slowly taken from me because for many businesses, agencies, etc.. thats the only way they are now accessed.''}

\subsection{External Source \texorpdfstring{($N=58$)}{Lg}}
Some respondents included links to articles, websites, videos, and/or other Quora discussions for readers seeking more information. Some examples include \textit{``Here is a link that will give you more insights on online privacy''} and \textit{``If you want to secure your android mobile data. See this video below, hope it may help you to secure your data.''} These external sources provide further content or support for the discussion question. Sometimes, representative of organizations or companies responded to users' queries and linked to supplementary information on their websites. For instance, a representative of the mobile network company, T-Mobile, responded to a question regarding its modified privacy policy: \textit{``We may occasionally ask you to confirm your choices. Please see our advertising choices section below for more information.''} 

\subsection{Personal Experience \texorpdfstring{($N=56$)}{Lg}}
Users also recounted their own experiences, preferences and actions in their responses, either as examples to reinforce their viewpoint or as a source of guidance. For instance, while arguing about the prevalence of Google in our every daily lives, one user wrote: \textit{``Let me tell you an interesting fact, I myself am using Quora on google right now. But that doesn't mean I'm entirely reliant on Google. I use Firefox often, I use VPN all the time and for my smartphone, I use this app Tor browser private web + vpn. I try to take at least some measures that can reassure that I'm trying to save my privacy.''} Users also often gave recommendations of apps or services that they personally used and tested, as this response shows: \textit{``I use LastPass . It is great and secure. If you are looking for open-source then KeePass (software) is not bad. Then you have 1Password , Dashlane and others,  I  tried them but they were not as good as Lastpass.''}

\subsection{Relationship Between Response Characteristics and Response Topics}
\begin{figure}[t]
    \centering
    \includegraphics[width=0.8\linewidth]{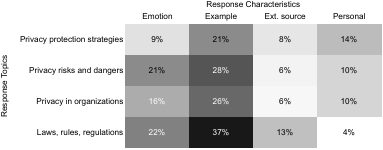}
    \caption{Percentages of responses containing each privacy topic identified with each response characteristic.}
    \label{fig:relations3}
\end{figure}

To investigate how the answers concerning each topic are phrased, we calculated the frequency of the response characteristics that appeared in the responses for each privacy topic (see Fig. \ref{fig:relations3}). Again, the percentages were calculated as a ratio of the total number of times a question type appeared with a particular privacy topic in the question against the total number of occurrences of that privacy topic in the questions. Examples were consistently the most abundant component found across all the response topics. Additionally, users frequently expressed emotions in responses touching on privacy laws, rules and regulations (22\%), and privacy risks and dangers (21\%). Personal experiences were also fairly frequently discussed in topics (more than 10\% of the responses) other than privacy laws, rules, and regulations. On the other hand, the external source was the least frequently seen characteristic in responses, except for those touching on privacy laws, rules, and regulations (13\%).

\section{Discussion} 
Our results revealed that online discussion platforms like Quora provided a collective sense-making space in which users explore the complexity of privacy together. Users were interested in a wide range of privacy topics, seeking varied types of information, and leveraging different resources to back up their discussions. Collectively, these discussions constructed a community-level mental model of user-perceived privacy. Our analysis also revealed that privacy topics included in the responses on Quora not only echoed those in the questions but also frequently touched upon other topics. These results indicated the importance of considering the bottom-up approach for improving users' privacy-protection attitude and competence and eventually promoting user agency. In this section, we synthesize our findings and discuss their implications for future directions of community-driven platforms for empowering users in privacy management.

\subsection{Synthesis of Results: A Four-Layer View of Users' Privacy Attitudes}
Our results indicated that user discussions span across four layers of factors related to their privacy concerns. At the \textbf{individual layer}, users discussed the problem of individuals being \textit{Ignorance About Privacy}, the importance of \textit{Raising Awareness}, and practicing \textit{Caution in Online Behavior}. At the \textbf{technology layer}, users worried about risks like \textit{Information Leaking or Hacking} and \textit{Identification and Deanonymization}; at the same time, they listed various technical solutions for protecting privacy, such as \textit{VPN and Firewall}, \textit{Encryption}, and proper \textit{Authentication}. At the \textbf{organization and application layer}, users raised concerns on practices of \textit{Third-Party Data Sharing} and \textit{Sale and Misuse of Data}, and they questioned \textit{Company's Privacy Attitudes and Data Handling Practices}; they advocated for \textit{Applications with Private and Secure Features} and \textit{Applications Built for Privacy and Security}, preferring apps that allow \textit{Adjusting Settings and Permissions}. Finally, at the \textbf{society layer}, users actively sought an understanding of the privacy \textit{Requirements Specified by the Regulatory Authorities}; they highlighted the trend of ubiquitous \textit{Tracking and Surveillance} and an overall \textit{Lack of Privacy on the Internet}. All four layers collectively affect users' attitudes toward privacy, which eventually influences the users' preferences and behaviors when using various technologies, applications, and platforms.

% This four-layer view of factors influencing privacy attitude discussed above can contextualize previous work and highlight future research directions. 
Most previous literature addressed a single layer of the factors given their study context. For example, studies focusing on redesigning the informational interface of apps' privacy practices to improve users' awareness and privacy protection behavior~\cite{Shiri2024, Kelley2010} addressed the individual layer. Similarly, the notion of ``Privacy by Design''~\cite{Wong2019} aimed at proposing best practices for making privacy-conscious services mostly focused on the organization and application layer. Although useful, these disjointed efforts fall short of adequately addressing privacy issues, as privacy is becoming increasingly entangled with every aspect of users' digital activities and society as a whole. This can be seen in examples like the privacy labels prevalently required by the most popular app stores but frequently deviate from the application's privacy practices, or the overnight adoption of cookie preferences by most online services to satisfy regulation requirements but the users' dismissive behavior using them.

Our results suggested that the users' perceptions of privacy are often intertwined, touching on multiple factors across individual, technological, organizational, and societal levels. This realization calls for \textbf{more holistic approaches to examining privacy, especially ones that cut across the intersections between different layers}. For example, we should investigate how to simultaneously encourage individuals to attend to their privacy and motivate application developers to incorporate best practices related to privacy (e.g., through building a community dedicated to distributed critique on privacy practice~\cite{KouG17}). Similarly, future efforts could focus on understanding the interplay among privacy-related expectations at the societal level (e.g., privacy literacy, the growing ubiquity of technology adoption, and the corresponding tracking and surveillance) and the development, adoption, and abandonment of privacy technologies.

\subsection{Fostering Community-Driven Approach of Privacy Support}
Previous studies have often focused on top-down interventions to address users' privacy concerns, through investigating the roles of regulations, standards, and technical solutions. While valuable, these approaches tended to overlook the importance of user agency and the crucial role of the user community. In contrast, our study took a different approach and focused on understanding the bottom-up efforts users made to collectively navigate the increasingly complex issue of privacy. Through analysis of users' existing behavior sharing and deliberating privacy-related topics on Quora, our results provided several implications for the design of future community-driven platforms to support collective sense making and community engagement.

\subsubsection{Supporting diverse information seeking and sharing behaviors}
Our analysis uncovered three main types of questions when users discuss privacy: opinion, factual, and guidance. Each type of questions represents a unique need for privacy-related information seeking and sharing that can be better accommodated with different community-based features. The \textit{opinion questions} seek highly subjective responses and sometimes touch on controversial topics. The goal of asking these questions is often to gather different perspectives, understand the community's tendencies, or challenge the status quo. In this scenario, features like quoting other responses, synthesizing multiple discussion threads, and automatically summarizing opposing points of view would help develop a more comprehensive view of the focal point and encourage continuous discussion. On the other hand, people who ask \textit{factual questions} often aim to acquire objective information about privacy. So supporting cross-referencing external resources for fact-checking and allowing the community to identify more accurate answers would be more beneficial~\cite{HoqueJMC17}. \textit{Guidance questions} fall in the middle; the quality and credibility of the answer depend on the experience and background of the person who provides it. Also, some advice is not suitable for all. The answerers' individual context can help individuals assess the applicability of the guidance provided in the response. So, ways to facilitate the answerers in providing these types of information would be useful. In general, community-based platforms should consider allowing users to specify the type of their privacy-related questions and provide corresponding features to promote community-driven knowledge sharing.

\subsubsection{Catering to users' diverse levels of technical expertise and different preferences}
Our results indicated that users mentioned a wide range of solutions, ranging from entirely technical approaches (e.g., VPN, firewall, and encryption), fine-tuned adjustments (e.g., adjusting settings and permissions), off-the-shelf solutions (e.g., applications with private features or built particularly for privacy), to individual behavioral change (e.g., raising awareness and exercising caution). This reflects the diverse levels of technical expertise of the user base on online communities like Quora. Moreover, when discussing \textit{privacy protection strategies}, users tended to mention one-stop and ready-to-use solutions more often than fine-tuned options that require more personal efforts. While control over individual apps is important (as discussed in previous studies such as by \citet{FengYS21}), users' self-agency diminishes with the increased number of apps and services they now use~\cite{solove}. Users often are not equipped with sufficient knowledge, are not provided with appropriate options, or simply do not have the time. These contexts should be incorporated into community-driven platforms. For example, these platforms can encourage and facilitate the creation of high-quality content that caters to diverse levels of technical expertise and preferences of users and make such content easily discoverable, such as through tagging or labeling with indicators of technical complexity. The platforms can also leverage their community moderation tools to support the curation of multiple answers with varying levels of complexity, allowing users to select or filter them based on their needs. Moreover, indicators of the technical expertise and preference of the questions can also be collected and displayed, so that other community members can provide more targeted answers.

\subsubsection{Simplifying the access to collectively-generated privacy insights}
Features that organize privacy-related insights from community discussions can benefit users and encourage their contribution to the collective sense-making process. Although Quora has features to organize a collection of questions (i.e., ``topic'' on Quora) and to organize a community around a specific topic (i.e., ``space'' on Quora), there is no dedicated topic nor space for privacy-related concerns, resulting in scattered information across the platform. For individual users, the discussions are also obscured by opaque and sometimes unreliable recommendation mechanisms. Thus, providing a coherent space for discussing privacy-related issues could foster informed dialogue and sense-making among users. Organizing related insights based on more detailed privacy topics, such as the ones identified in this work, can better orient users. Summarizing those insights, possibly leveraging previous CSCW work through community-driven approaches (e.g., Wikum by \citet{zhang2017wikum}) or combining semi-automated techniques (e.g., SUMMIT by \citet{Gilmer2023SUMMIT}), can further synthesize diverse user opinions and drive for more in-depth discussion.

\subsubsection{Creating safe spaces for discussing tradeoffs and personal contexts} 
As a result of the complexity of privacy management, many users expressed a tendency to trade privacy for the convenience offered by the technology. Users also frequently voiced concerns that big companies generally operate on the ``economics of personal data''~\cite{10.1093/idpl/ipw026}, while acknowledging that it is nearly impossible to avoid using their applications and services in the current technology landscape. This is supported by previous research indicating a privacy/convenience tradeoff experienced by users~\cite{10.1145/2470654.2481328, WEINBERG2015615}, as well as aspects of the privacy calculus theory~\cite{laufer1977privacy}, which posits that individuals engage in cost-benefit analysis when deciding to disclose personal information, weighing the perceived benefits against potential privacy risks. However, this type of analysis is highly influenced by the personal contexts of the users, which can change over time. Indeed, as demonstrated in our results, privacy was frequently discussed as a subjective experience, laden with personal stories and emotions. Users coped with the complexity and sought tradeoffs by adopting strategies tailored to their individual situations and needs. Thus, future work should investigate the scope of contextual factors that may influence users' privacy-related decision-making process and incorporate them into online community platforms. Examples of such solutions may include collecting user preferences or learning user behavior before suggesting privacy best practices or community-generated content. It worth acknowledging that approaches like this have their own privacy concerns. Balancing those concerns with user values is a constant challenge in online communities that requires continuous attention.

\subsubsection{Connecting technology designers and developers with the user community} 
Empowering users also relies on improving designers' and developers' empathy and understanding of users' privacy-related struggles. Previous studies have found that, when it came to privacy, developers predominantly engaged in discussions that revolve around different phases of application development~\cite{10.1145/3432919}, encompassing programming practices and legal considerations pertaining to privacy requirements~\cite{07daf61b7df048c188e185079294a59c, 10.1145/3313831.3376768}. Developers possess the technical prowess to address privacy issues~\cite{10.1145/3432919} and their decisions directly impact users~\cite{DBLP:conf/chi/TahaeiAR23}. However, the subjective and non-technical aspect of privacy, which is extensively discussed in our data, rarely appears in developer-focused forums and may not be well-recognized by designers and developers. Future work should thus focus on helping them establish empathy toward users' concerns and frustration with privacy issues. One possible approach is through aggregating privacy-related feedback, from user forums, app reviews, and question-answering platforms, to help designers and developers better understand users' struggles. More broadly, bringing together the previously separate communities of designers, developers, and users can go a long way to fostering a deeper understanding of users' struggles and creating more effective and user-centered privacy solutions.

\subsection{Limitations}
We acknowledge several limitations related to our methods. First, we only focused on Quora discussions as our main data source. Although this platform provides unique benefits to understanding users' collective sense-making approach, users' privacy-related discussions on other platforms may reveal additional aspects that require future work to explore. Second, we used specific keywords to search Quora questions during our data collection process. Although the keywords were carefully selected, we can not claim that they are exhaustive. We might still have left out some discussion topics that were relevant to our study and these discussions or topics could have altered our findings. Third, our final analysis was performed on a sample of the dataset. Although the sample size was calculated to represent the population and the sampling was performed randomly, a sampling error could still occur and bias our results. Finally, our analysis was influenced by our experiences and positionality. All authors are experienced in HCI and software engineering and have investigated user privacy in prior projects. Researchers with different backgrounds may approach the analysis differently.

\section{Conclusion}
The importance of information privacy is more critical than ever in the current digital landscape, especially for technology users. Through analyzing the discussions about privacy-related topics on Quora, we identified four categories of topics users touched on in their questions and responses. Among them, privacy topics such as privacy protection strategies, particularly regarding secure applications, and privacy risks and dangers, including tracking and information leaks or hacking, were most discussed. Meanwhile, our analysis suggests that privacy-related questions frequently seek personal opinions, aside from factual information and guidance. At the same time, responses to different topics are rich in information, such as examples and external links, and reflect personal experience and values. Overall, our study depicts the complexity of users' perspectives concerning privacy, calling for more holistic approaches that connect the individual, technological, organizational, and societal layers that affect users' privacy attitudes. Moreover, we suggest directions moving forward to better foster users' community-driven sense-making of privacy and empower users to navigate their subjective, personal experience managing privacy-related issues.

\begin{acks}
This research is partially supported by the Fonds de recherche du Québec (2022-PR-300101) and the Canada Research Chairs program (CRC-2021-00076).
\end{acks}

\bibliographystyle{ACM-Reference-Format}
\bibliography{reference}

% \appendix

\end{document}